\documentclass[graybox]{svmult}

\usepackage{mathptmx}       
\usepackage{helvet}         
\usepackage{courier}        
\usepackage{type1cm}        
\usepackage{makeidx}         
\usepackage{graphicx}        
\usepackage{multicol}        
\usepackage[bottom]{footmisc}

\usepackage{amsmath}

\makeindex             

\usepackage{todonotes}

\begin{document}

\title*{Fast variables determine the epidemic threshold in the pairwise model with an improved closure}
\author{Istv\'an Z. Kiss, Joel C. Miller and P\'eter L. Simon}
\institute{Istv\'an Z. Kiss \at School of Mathematical and Physical Sciences, Department of
Mathematics, University of Sussex, Falmer,
Brighton BN1 9QH, UK, \email{i.z.kiss@sussex.ac.uk}
\and Joel C. Miller \at Institute for Disease Modeling, Bellevue, WA,
United States of America \email{joel.c.miller.research@gmail.com}
\and P\'eter L. Simon \at Institute of Mathematics, E\"otv\"os Lor\'and University Budapest, and
Numerical Analysis and\\
Large Networks Research Group, Hungarian Academy of Sciences, Hungary,
\email{simonp@math.elte.hu}}
%
%
\maketitle


\abstract*{Pairwise models are used widely to model epidemic spread on networks.
These include the modelling of susceptible-infected-removed (SIR) epidemics 
on regular networks and extensions to
SIS dynamics and contact tracing on more exotic networks exhibiting degree heterogeneity, 
directed and/or weighted links and clustering.
However, extra features of the disease dynamics or of the network lead to an increase in  system size 
and analytical tractability becomes problematic.  Various ``closures''
can be used to keep the system tractable.  Focusing on SIR epidemics on regular but clustered networks, we show
that even for the most complex closure we can determine the epidemic threshold as an asymptotic expansion in terms of the 
clustering coefficient.  We do this by exploiting the presence of a system of fast variables, 
specified by the correlation structure of the epidemic, whose 
steady state determines the epidemic threshold. While we do not find
the steady state analytically, we create an elegant asymptotic expansion of it.
We validate this new threshold by comparing it to the numerical
solution of the full system and find excellent agreement over a wide range of values of the clustering 
coefficient, transmission rate and average degree of the network. The technique carries over to pairwise models with other closures~\cite{barnard2018epidemic} and we note that the 
epidemic threshold will be model dependent. This emphasises the importance of model choice when dealing with realistic outbreaks. 
 }

\abstract{Pairwise models are used widely to model epidemic spread on networks.
These include the modelling of susceptible-infected-removed (SIR) epidemics 
on regular networks and extensions to
SIS dynamics and contact tracing on more exotic networks exhibiting degree heterogeneity, 
directed and/or weighted links and clustering.
However, extra features of the disease dynamics or of the network lead to an increase in  system size 
and analytical tractability becomes problematic.  Various ``closures''
can be used to keep the system tractable.  Focusing on SIR epidemics on regular but clustered networks, we show
that even for the most complex closure we can determine the epidemic threshold as an asymptotic expansion in terms of the 
clustering coefficient.  We do this by exploiting the presence of a system of fast variables, 
specified by the correlation structure of the epidemic, whose 
steady state determines the epidemic threshold. While we do not find
the steady state analytically, we create an elegant asymptotic expansion of it.
We validate this new threshold by comparing it to the numerical
solution of the full system and find excellent agreement over a wide range of values of the clustering 
coefficient, transmission rate and average degree of the network. The technique carries over to pairwise models with other closures~\cite{barnard2018epidemic} and we note that the 
epidemic threshold will be model dependent. This emphasises the importance of model choice when dealing with realistic outbreaks. 
}

\section{Introduction} \label{Intro}
\label{sec:1}
One way to deal with the challenges of modelling stochastic epidemics on networks is to use mean-field models. This approach has led to a number of models including heterogeneous or degree-based mean-field~\cite{pastor2001epidemic,pastor2015epidemic}, pairwise~\cite{rand1999correlation,keeling1999effects}, effective-degree~\cite{lindquist2011effective}, edge-based compartmental~\cite{miller2012edge} and message passing~\cite{karrer2010message}, to name a few. The main difference between these models is how the variables over which averaging is done are chosen. Perhaps the most compact model is the edge-based compartmental model~\cite{miller2013model} and this works for heterogeneous networks with Markovian SIR epidemics, although extensions of it for arbitrary infection and recovery processes are also possible~\cite{sherborne2018mean}.

Pairwise models are popular and the first model for regular networks
and SIR epidemics~\cite{rand1999correlation,keeling1999effects} was
generalised to heterogeneous networks~\cite{eames2002modeling},
preferentially mixing networks~\cite{eames2002modeling},
directed~\cite{sharkey2006pair} and weighted
networks~\cite{rattana2013class}, adaptive
networks~\cite{kiss2017mathematics}, and structured
networks~\cite{house2009motif} among others. Its wide use is
perhaps due to its relative transparency where variables are defined
in a straightforward way. A downside of the pairwise models is that in
constructing them we find that the change in the expected number of individual nodes of a
given state depends on to the expected number of edges (or pairs) between
nodes of various states.  The change in the expected number of edges depends
on larger-scale structure.  To keep the system tractable, we generally
make a ``closure assumption'' that we can express the frequency of the relevant
larger-scale structures in terms of the pairs and individuals, that is lower order moments or structure.

A basic understanding of the network and epidemic dynamics coupled
with careful bookkeeping and an appropriate closure assumption produces
a pairwise model. Pairwise models have been successfully used to
analytically derive the epidemic threshold and final epidemic
size. However, these results are mostly limited to networks without
clustering. The propensity of contacts to cluster, i.e. that two
friends of an individual/node are also friends of each other, is known
to lead to many complications, and modelling epidemics on clustered
networks using analytically tractable mean-field models is still
limited to networks with very specific structural
features~\cite{house2009motif,newman2009random,miller2009percolation,miller2009spread,karrer2010random,volz2011effects,ritchie2016beyond}. However,
using approaches borrowed from percolation
theory~\cite{miller2009spread} and focusing more on the stochastic
process itself~\cite{trapman2007analytical}, some results have been
obtained. 

For pairwise models, clustering first manifests itself by requiring a
different and more complex closure, which makes the analysis of the
resulting system, even for regular networks and SIR dynamics,
challenging. Furthermore, it turns out that such closures may in fact
fail to conserve pair-level relations and may not accurately reflect
the early growth of quantities such as closed loops of three nodes
with all nodes being infected~\cite{house2010impact}. Such
considerations have led to an improved closure being developed in an
effort to keep as many true features of the exact epidemic process as
possible~\cite{house2010impact}. In this paper we will focus on the
classic pairwise model for regular networks with clustering but using
the improved closure of~\cite{house2010impact}, given below in equation~\eqref{eq:improved_closure}.  
We will show that by working with fast variables corresponding to the correlations that develop during the spread of the epidemic, we can analytically determine the epidemic threshold as an asymptotic expansion in terms of the clustering coefficient. 

The use of fast variables is not completely new.  They were used
in~\cite{keeling1999effects} and~\cite{eames2008modelling} but not with the improved closure. 
Even with the simpler closures, the epidemic threshold has only been obtained numerically and it was
framed in terms of a growth-rate-based threshold (which is equivalent
to the basic reproduction number at the critical point of the epidemic
spread). In~\cite{eames2008modelling} a hybrid pairwise model
incorporating random and clustered contacts is considered, with the
analysis focused on the growth-rate-based threshold. The authors of
\cite{eames2008modelling} managed to derive a number of results, some
analytic (the critical clustering coefficient for which an epidemic
can take off) and some semi-analytic, and they have shown, in
agreement with most studies, that clustering inhibits the spread of
the epidemic when compared to an equivalent network without clustering
but with equivalent parameter values governing the epidemic
process. However, no analytic expression for the threshold was
provided. More recently, in~\cite{li2018epidemic}, the epidemic
threshold in a pairwise model for clustered networks with closures
based on the number of links in a motif, rather than nodes, was
calculated. 

Building on these results and the recent paper by Barnard et al~\cite{barnard2018epidemic} (where the idea of fast variables was used to derive and analytic epidemic threshold for pairwise models with two different closures corresponding to clustered networks) 
we set out to take the final step of using fast variables and
perturbation theory to determine an asymptotic expansion of the
epidemic threshold when the pairwise model is closed with
equation~\eqref{eq:improved_closure}. The paper is structured as
follows. In Section 2 we outline the model. The main results, both
analytical and numerical, are presented in Section 3.  We conclude
with a discussion of the results and possible extensions in Section 4.

\section{Model formulation} \label{ModelForm}

\subsection{The network and standard SIR dynamics} \label{ModelForm_network}

We begin by considering a population of $N$ individuals and describe their
contact structure by an undirected network with adjacency
matrix $G=(g_{ij})_{i,j=1,2,\dots, N}$ where $g_{ij}=1$ if nodes $i$
and $j$ are connected and zero otherwise.  Because the network is
undirected, $g_{ij}=g_{ji}$ for all $i,j=1,2, \dots N$, and because we
exclude self-loops, $g_{ii}=0$ for all $i$. The network is static and regular, such that each individual has exactly $n$ edges or links. The sum over all elements of $G$ is defined as $||G||=\sum_{i,j}g_{ij}$. Hence, the number of doubly counted links in the network is $||G||=nN$. More importantly, using simple matrix operations on $G$, we can calculate the clustering coefficient of the network 
\begin{equation} \label{equation:phi_clustering}
\phi=\frac{trace(G^{3})}{||G^{2}||-trace(G^{2})}, 
\end{equation}
where $trace(G^{3})$ yields six times the number of closed triples or loops of length three (uniquely counted) and $||G^{2}||-trace(G^{2})$, twice the number of triples (open and closed, also uniquely counted).

Let us consider the standard SIR epidemic dynamics on a network. The
dynamics are driven by two processes: (a) infection and (b) recovery
from infection. Infection can spread from an infected/infectious node
to any of its susceptible neighbours.  We model this as a Poisson
point process with per-link infection rate $\tau$. Infectious nodes
recover at constant rate $\gamma$, independently of the network, and
gain permanent immunity.

\subsection{The unclosed pairwise model} \label{ModelForm_UnClosedPWModel}
Let $A_{i}$ be 1 if the individual at node $i$ is of type $A\in \{S, I, R\}$ and zero otherwise. Then single nodes (singles) of type $A$ can be counted as $[A]=\sum_{i}A_{i}$, pairs of nodes (pairs) of type $A-B$ can be counted as $[AB]=\sum_{i,j}A_{i}B_{j}g_{ij}$ and triples of nodes (triples) of type $A-B-C$ can be counted as $[ABC]=\sum_{i,j,k}A_{i}B_{j}C_{k}g_{ij}g_{jk}$. This method of counting means that pairs are counted once in each direction, so $[AB]=[BA]$, and $[AA]$ is even. Using this notation to track singles, pairs, and triples leads to the following system of pairwise equations describing the SIR epidemic on a regular network:
\begin{align} 
\dot{[S]}&=-\tau[SI]; \,\,\,\,\,\,\, \dot{[I]}=\tau[SI] -\gamma[I]; \,\,\,\,\,\,\,\dot{[R]}= \gamma[I],\label{equations:unclosed_pw_model_one} \\ 
\dot{[SI]}&=\tau([SSI]-[ISI]-[SI])-\gamma[SI], \label{equations:unclosed_pw_model_two}\\
\dot{[SR]}&=\gamma [SI]-\tau[RSI], \label{equations:unclosed_pw_model_three} \\ 
\dot{[II]}&=2\tau([ISI]+[SI])-2\gamma[II], \label{equations:unclosed_pw_model_four} \\
\dot{[IR]}&=\gamma([II]-[IR])+\tau[RSI].\label{equations:unclosed_pw_model_end}
\end{align}
 We note that equations \eqref{equations:unclosed_pw_model_two}--\eqref{equations:unclosed_pw_model_end} contain triples which are not defined within the entire system of equations \eqref{equations:unclosed_pw_model_one}--\eqref{equations:unclosed_pw_model_end}. Furthermore, we have chose these variables in order to be able to 
 consistently define our fast variables later.
To determine solutions of the system, we must find a way to account
for these triples in terms of pairs and singles through a closure assumption.
It is worth noting that this system is exact before a closure is
implemented~\cite{kiss2017mathematics}.

\subsection{The improved closure and the closed pairwise system}

The key for deriving the improved closure~\cite{house2010impact} is to
split the non-clustered and clustered part of the network and to
determine the propensity of a susceptible node's neighbour to be in state $A$ (where $A\in\{S,I,R\}$), given that the susceptible node is already connected to an infected one. 
This can be defined as
\begin{equation}
p_{A|S-I}=   \begin{cases}
     p_{A|S-I}^{uc} & \text{with probability}\ (1-\phi),\\
     \frac{p_{A|S-I}^{c}}{\sum_{a}p_{a|S-I}^{c}} & \text{with probability}\ \phi,
    \end{cases}
\label{eq:clo_clust_improved}
\end{equation}
where $p_{A|S-I}^{uc}=\frac{[AS]}{n[S]}$,
$p_{A|S-I}^{c}=p_{A|S-I}^{uc}C_{AI}$ and
$C_{AI}=\frac{N[AI]}{n[A][I]}$.  In the absence of clustering we assume that 
the probability the neighbour is of state $A$ is simply given by
frequency of $[AS]$ type links relative to all links emanating from
susceptible nodes, $n[S]$. If clustering is present then the probability
of finding a susceptible neighbour decreases as the transitive link
connects this particular neighbour to the existing infected
neighbour. This means that the node is exposed to infection and its
probability of remaining susceptible decreases. This effect is
captured by $C_{AI}$ which expresses how much more probable it is,
compared to the random mixing case, to find a neighbour in state $A$
given that the 
 node is also connected to an infectious node. It is well know that epidemics are negatively correlated in the sense that we are more likely to find
$I-I$ type links rather than $I-S$.  Unfortunately, $p_{A|S-I}^{c}$
alone is not a properly defined probability. Despite this the closure
resulting from it has been used although it leads to some anomalies
such as non-conservation of pair-level relations. However, the
normalised form of it, as in equation~\eqref{eq:clo_clust_improved},
leads to the improved closure~\cite{house2010impact}. Taking into account the new way of defining $p_{A|S-I}$, this yields
\begin{align}
[ASI]_c&=(1-\phi)[ASI]+\phi[ASI]= (1-\phi)(n-1)[SI]p_{A|S-I}^{uc}+\phi (n-1)[SI]\frac{p_{A|S-I}^{c}}{\sum_{a}p_{a|S-I}^{c}}\notag\\
&= (1-\phi)(n-1)[SI]\frac{[AS]}{n[S]}+\phi (n-1)[SI]\frac{\frac{[AS]}{n[S]}C_{AI}}{\sum_{a}p_{a|S-I}^{c}}\notag\\
&= (1-\phi)\frac{(n-1)}{n}\frac{[AS][SI]}{[S]}+\phi (n-1)[SI]\frac{\frac{[AS]}{n[S]}\frac{N[AI]}{n[A][I]}}{\sum_{a}\frac{[aS]}{n[S]}\frac{N[aI]}{n[a][I]}}\notag\\
&=(1-\phi)\frac{(n-1)}{n}\frac{[AS][SI]}{[S]}+\phi (n-1)\frac{[AS][SI][IA]}{[A]\sum_{a}\frac{[aS][aI]}{[a]}}\notag\\
&=(n-1)\left((1-\phi)\frac{[AS][SI]}{n[S]}+\phi \frac{[AS][SI][IA]}{[A]\sum_{a}[aS][aI]/[a]}\right),
\label{eq:improved_closure}
\end{align}
where $a\in\{S,I,R\}$ and $[ASI]_c$ is used to distinguish this approximation from its exact equivalent.

\section{Results for the pairwise model with the improved closure} \label{section:Results_comp_imp}
Plugging equation \eqref{eq:improved_closure} into the exact system~\eqref{equations:unclosed_pw_model_one}--\eqref{equations:unclosed_pw_model_end} leads to the self-consistent system below
\begin{align} 
\dot{[S]}&=-\tau[SI]; \,\,\,\,\,\,\, \dot{[I]}=\tau[SI]-\gamma[I]; \,\,\,\,\,\,\, \dot{[R]}=\gamma[I]\label{eq:dot_S_I_R_imp_close}\\
\dot{[SI]}&=-(\tau+\gamma)[SI]+\tau[SSI]_c-\tau[ISI]_c, \label{eq:dot_SI_reduced_imp_close}\\
 \dot{[SR]}&=\gamma[SI]-\tau[RSI]_c, \label{eq:dot_SR_imp_close}\\
 \dot{[II]}&=2\tau[SI]-2\gamma[II]+2\tau[ISI]_c, \label{eq:dot_II_imp_close}\\
 \dot{[IR]}&=\gamma([II]-[IR])+\tau[RSI]_c, \label{eq:dot_IR_imp_close}
 \end{align}
where $[ASI]_c$ with $A\in\{S,I,R\}$ is defined in equation~\eqref{eq:improved_closure}. The standard linear stability analysis of this system around the disease free steady state, $([S],[I],[SI],[SS],[II])=(N,0,0,nN,0)$ leads to 
some terms such as 
\begin{equation}
\alpha=\frac{[SI]}{[I]}, \delta=\frac{[II]}{[I]}, x=\frac{[SR]}{[R]}, y=\frac{[IR]}{[I]}.
\label{eq:def_fast_var}
\end{equation}
Interestingly these terms are ill-defined since both denominators and
numerators are zero at the equilibrium. However,
these variables have a clear biological meaning and are related to the correlation structure of the epidemic.

Interestingly however, the epidemic threshold can also be found in a more direct way by looking at equation~\eqref{eq:dot_S_I_R_imp_close}. Namely, this leads to
\begin{equation}
\dot{[I]}=\tau[SI]-\gamma[I]=\gamma[I]\left(\frac{\tau}{\gamma}\frac{[SI]}{[I]}-1\right),
\label{eq:growth_rate_based_threshold}
\end{equation}
which clearly shows that the epidemic threshold coincides with $\overline{\mathcal{R}}_0=\frac{\tau}{\gamma}\frac{[SI]}{[I]}=1$.
This is a growth-rate-based threshold of the epidemic and while
$\overline{\mathcal{R}}_0$ is different from the basic reproduction
number, they are equivalent when both are exactly one. From here, we can see that finding the threshold amounts to finding $\alpha=\frac{[SI]}{[I]}$ at time $t$ close to zero.
As we will show next, these new variables of interest are fast
variables and settle quickly, even if only temporarily, to a
quasi-equilibrium.  The time taken to reach this quasi-equilibrium is
short compared to the timescale of epidemic growth, and the quasi-equilibrium corresponds to the exponential growth phase of the epidemic.
\begin{figure}
\centering
{\includegraphics[scale=0.19]{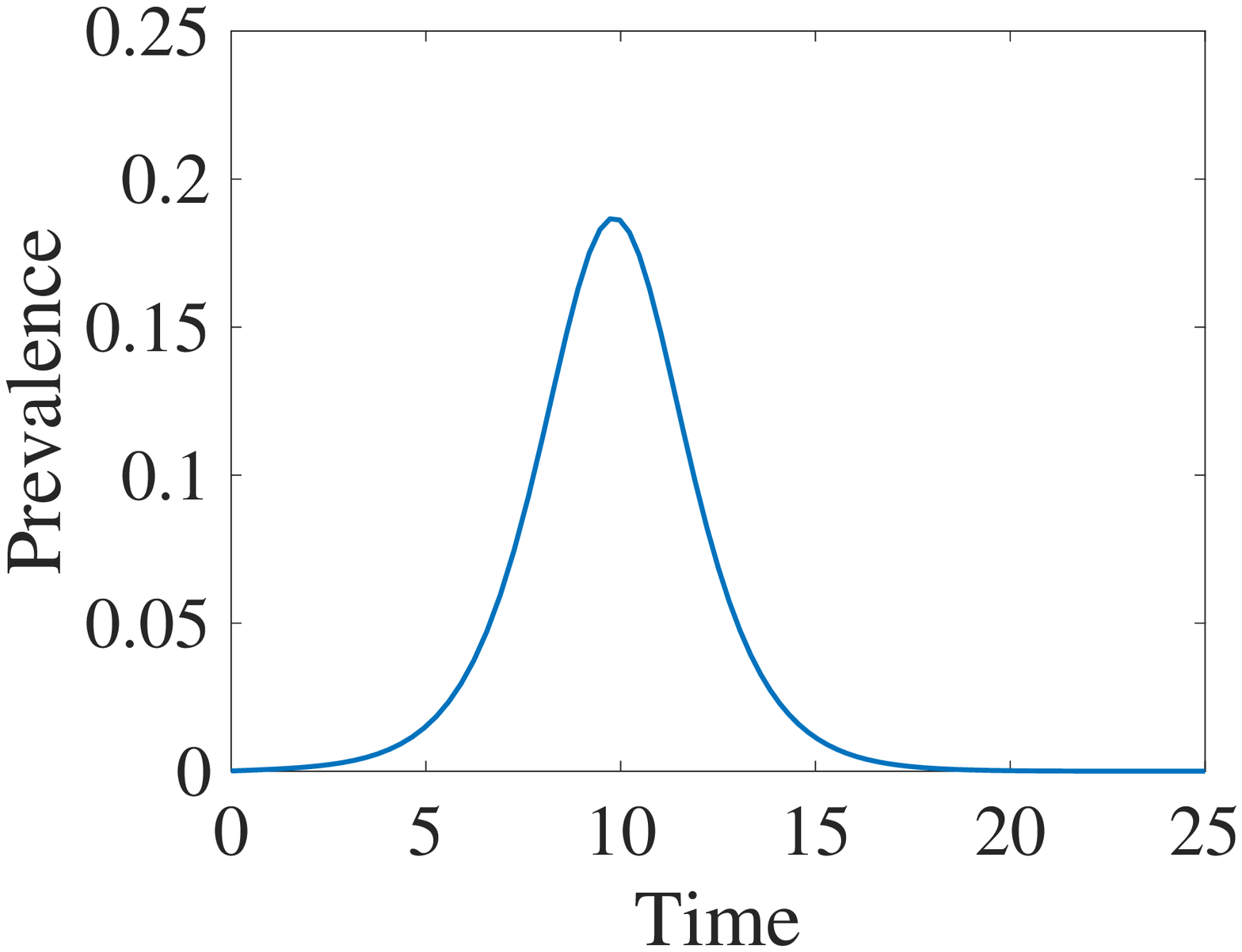}
    }
\hspace{0.1cm}
{\includegraphics[scale=0.19]{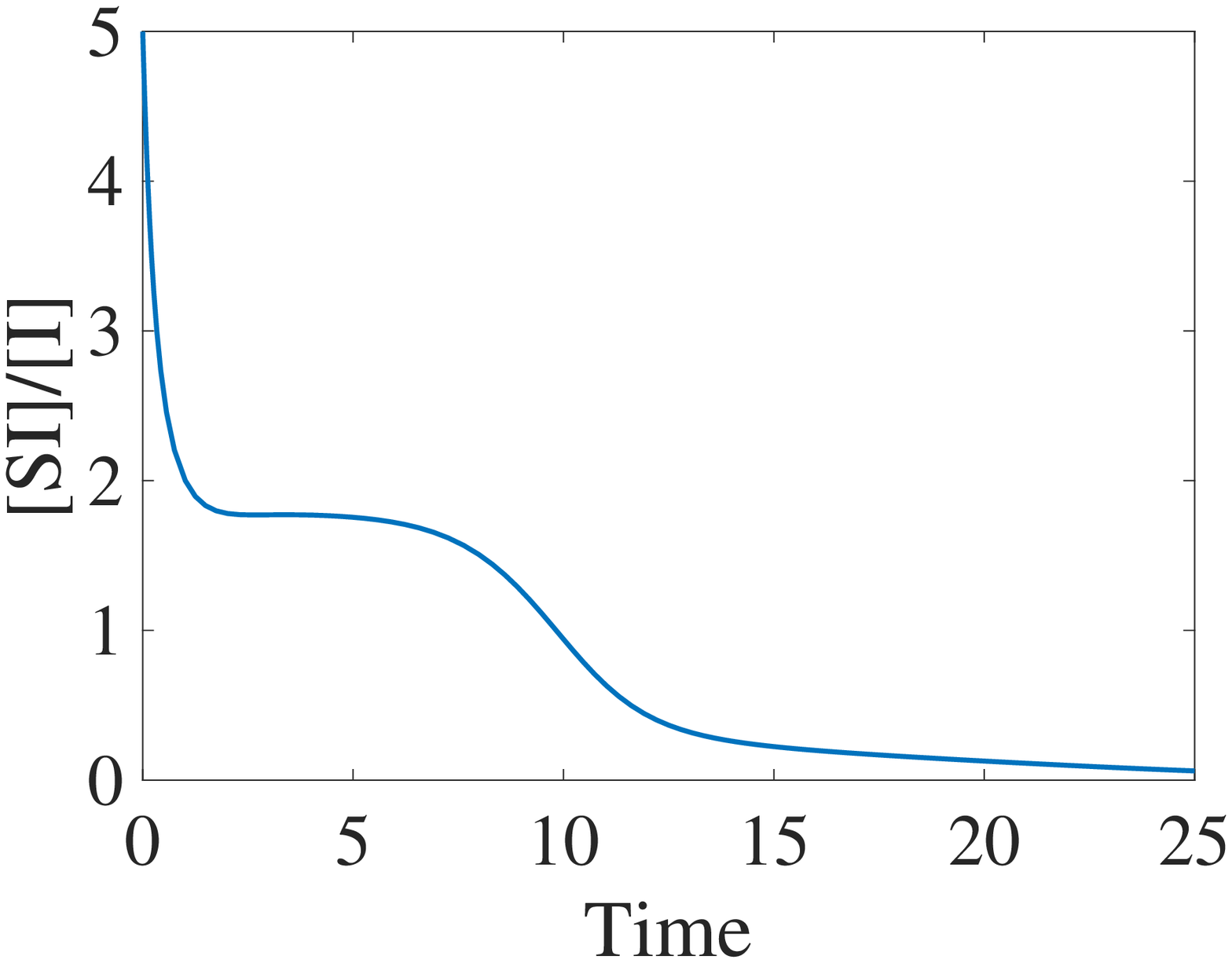}
    }
{\includegraphics[scale=0.19]{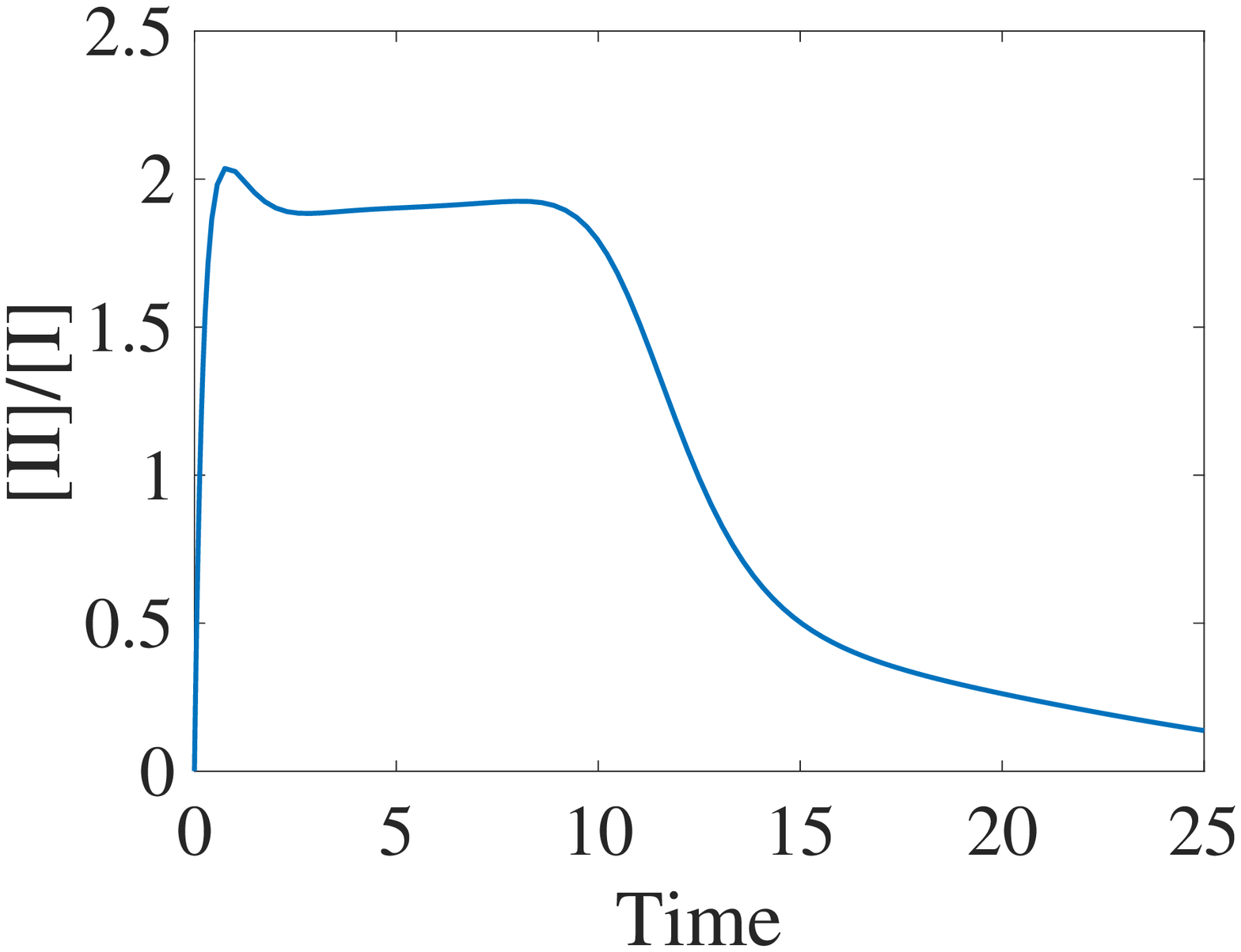}
        }
\hspace{0.1cm}
{\includegraphics[scale=0.19]{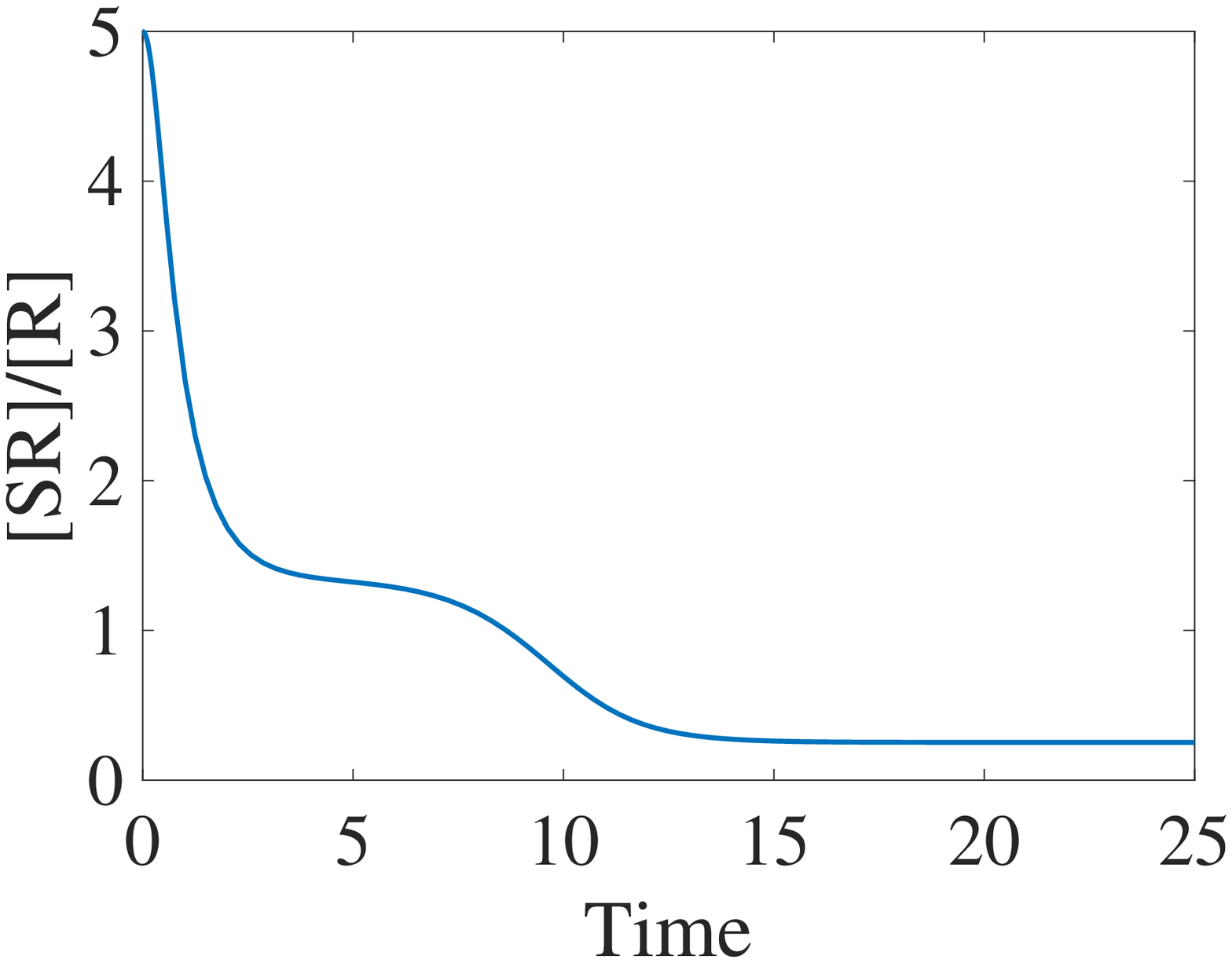}
  }
{\includegraphics[scale=0.19]{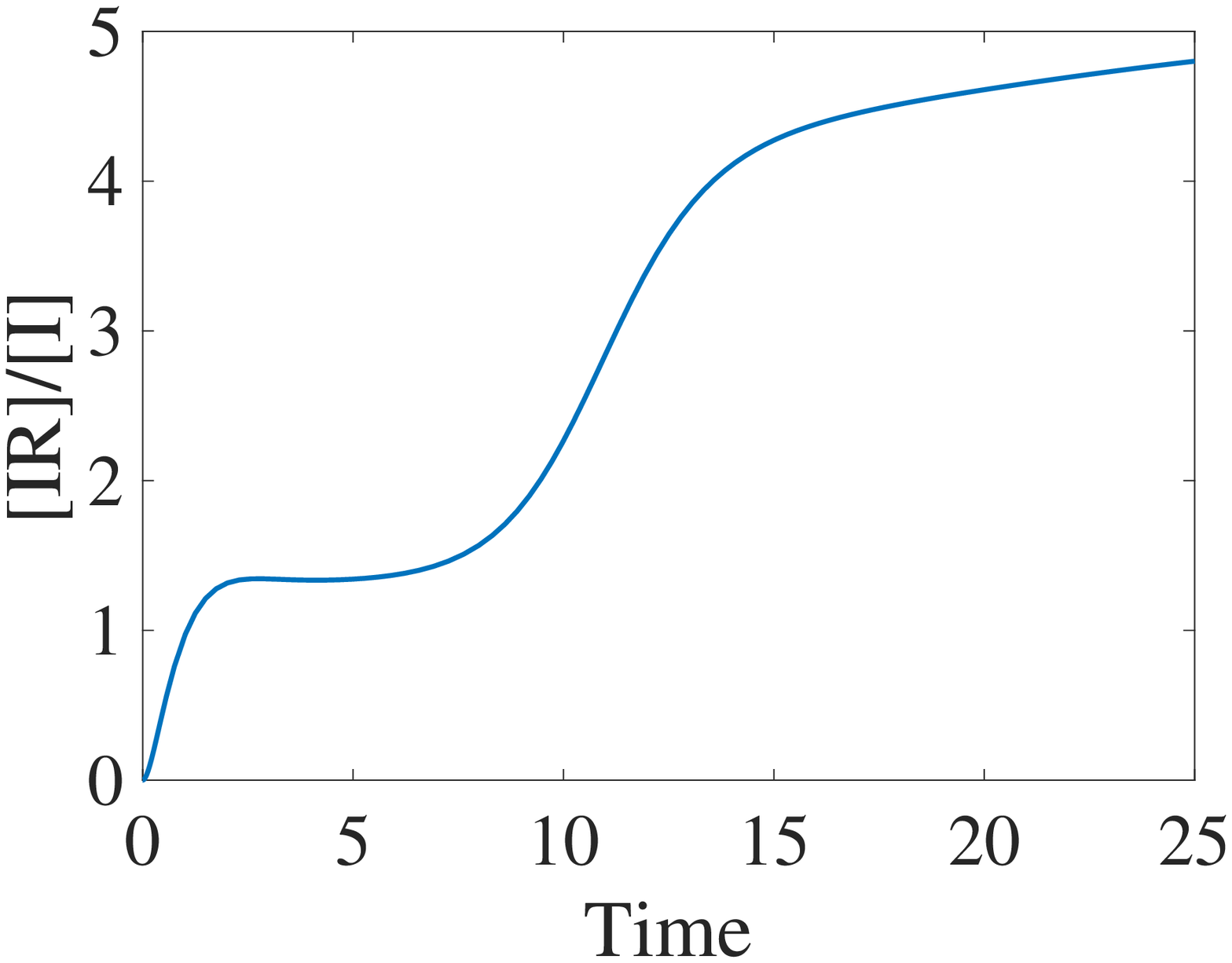}
        }
\caption{Time evolution of the prevalence and four fast variables
  based on numerical integration of the closed pairwise system
  \eqref{eq:dot_S_I_R_imp_close}--\eqref{eq:dot_IR_imp_close}. Parameter
  values are $N=10000$, \ $\gamma=1$ and $\phi=0.5$.}
  \label{fig:prev_and_fast_var}
\end{figure}
\subsection{Fast variables with the improved closure}
In Fig.~\ref{fig:prev_and_fast_var} we plot the evolution of the prevalence together with the newly defined variables. The figure shows clearly that these new variables are fast. Namely, early on, when the prevalence is small, these variables settle to a temporary equilibrium. The natural step would be to investigate the behaviour of $\alpha$ (and all the others) and this can be done by deriving their evolution equations, i.e. compute $d\alpha/dt=d([SI]/[I])/dt$ and use equations~\eqref{eq:dot_S_I_R_imp_close}--\eqref{eq:dot_IR_imp_close}. After some simple but lengthy algebra one obtains 
\begin{equation}
\frac{d\alpha}{dt}=-\tau \alpha(1+\alpha)+\tau \xi (1-\phi) n \alpha+\tau (n-1)\phi \left(\frac{n \alpha^2-\alpha^2\delta}{n\alpha+\alpha \delta+xy}\right)\label{eq:alpha}.
\end{equation}
The assumption of being close to $t=0$ is used to neglect a term of
the form $\tau \xi (1-\phi)\frac{[SI]}{[I]}\frac{[SI]}{[S]}$,
together with a few others with a similar structure.  While $\alpha=[SI]/[I]$
itself is a well-defined and bounded ratio of two small numbers,
$[SI]/[S]\simeq 0$ when $t$ is close to zero.  We use similar arguments when deriving the equations for the other variables. Their differential equations are
\begin{align}
\frac{d\delta}{dt}&=-\gamma \delta+2\tau\alpha-\tau \alpha \delta+2\tau (n-1)\phi \frac{\alpha^2\delta}{n\alpha+\alpha \delta+xy},\label{eq:delta}\\
\frac{dx}{dt}&=+\tau\alpha^{*}(\alpha-x)-\gamma(\alpha-x)-\frac{\tau(n-1)\phi \alpha x y\left(\frac{\tau}{\gamma}\alpha^{*}-1\right)}{n\alpha+\alpha \delta+xy},\label{eq:x}\\
\frac{dy}{dt}&=+\tau \delta-\tau \alpha y+\frac{\tau(n-1)\phi \alpha x y}{n\alpha+\alpha \delta+xy}.\label{eq:y}
\end{align}
As one notices the four variables are interlinked and are all needed
to resolve the evolution equation of each.  A key step in the derivation above is the need to introduce $\alpha^{*}$ which corresponds to the steady state of the system defined by equations~\eqref{eq:alpha}--\eqref{eq:y}. This is needed as in the derivation of the evolution equations for $x$ terms such as $[SI]/[R]$, $[IR]/[R]$ and $[I]/[R]$ can only be dealt with by noticing that at time $t$ close to $t=0$ we have that 
\begin{equation}
[I]=\left(\frac{\tau}{\gamma}\alpha^{*}-1\right)[R].
\end{equation} 
This follows from the assumption that $[I] \approx ce^{rt}$ for some $c$ and
$r$ and $d[R]/dt = \gamma I$, or from the observation that 
\begin{align}
\frac{d[I]}{d[R]}=\frac{\tau[SI]-\gamma [I]}{\gamma [I]}=\left(\frac{\tau}{\gamma}\frac{[SI]}{[I]}-1\right)\simeq \left(\frac{\tau}{\gamma}\alpha^{*}-1\right),
\end{align}
where we assumed that $\alpha$ stabilises quickly at small time. Integrating this leads to
$[I]=\left(\frac{\tau}{\gamma}\alpha^{*}-1\right)[R]+\text{C}$, where $C=0$ if the initial conditions at $t=0$ are plugged in. This in turn allows us to write 
$[SI]/[R]=([SI]/[I])([I]/[R])$ and $[IR]/[R]=([IR]/[I])/([I]/[R])$  which ensures that we can cast all terms as functions of the four fast variables.

\subsection{Asymptotic expansion of the epidemic threshold}
Finding the steady state of the system defined by equations~\eqref{eq:alpha}--\eqref{eq:y} may seem like a difficult task but it turns out that an asymptotic solution is within reach.
To do this each variable $v$ is written as $v=v_0+\phi v_1+\cdots$, where $v\in\{\alpha,\delta,x,y\}$. Plugging these into equations~\eqref{eq:alpha}--\eqref{eq:y} leads to the following system at $\mathcal{O}(1)$:
\begin{align}
((n-2)-\alpha_0)(n\alpha_0+\alpha_0\delta_0+x_0y_0)&=0,\\
(-\gamma\delta_0+2\tau\alpha_0-\tau\alpha_0\delta_0)(n\alpha_0+\alpha_0\delta_0+x_0y_0)&=0,\\
(\delta_0-\alpha_0y_0)(n\alpha_0+\alpha_0\delta_0+x_0y_0)&=0,\\
(\tau\gamma\alpha_0^2-\tau\gamma\alpha_0x_0-\gamma^2\alpha_0+\gamma^2x_0)(n\alpha_0+\alpha_0\delta_0+x_0y_0)&=0.
\end{align}
One of the solutions of the system above is:
\begin{equation}
\left(\alpha_0,\delta_0,x_0,y_0\right)=\left(n-2,\frac{2\tau(n-2)}{\gamma+\tau(n-2)},n-2,\frac{2\tau}{\gamma+\tau(n-2)}\right).\label{eq:Ord_one_sol}
\end{equation}
At $\mathcal{O} (\phi)$ from equation~\eqref{eq:alpha} we have
\begin{equation}
-(\alpha_1+(n-1))(n\alpha_0+\alpha_0\delta_0+x_0y_0)+(n-1)\alpha_0(n-\delta_0)=0.
\end{equation}
Plugging in the solutions at $\mathcal{O} (1)$ (see eq. \eqref{eq:Ord_one_sol}) into the equation above leads to 
\begin{equation}
\alpha_1=-\frac{2\tau (2n-3)(n-1)}{n(\gamma+\tau(n-2))+2\tau(n-1)}.
\end{equation}
Hence the epidemic threshold, up to the first correction is given by
$\overline{\mathcal{R}}_0=1$ where  
\begin{equation}
\overline{\mathcal{R}}_0=\frac{\tau}{\gamma}(\alpha_0+\phi \alpha_1)=\frac{\tau(n-2)}{\gamma}-\frac{\tau}{\gamma}\frac{2\tau (2n-3)(n-1)}{n(\gamma+\tau(n-2))+2\tau(n-1)}\phi.
\label{eq:threshold}
\end{equation}
The first observation that can be made is that the first order
correction is negative and this implies that clustering reduces the
epidemic threshold and makes the epidemic less likely to spread. The
second is that when $\phi=0$, \ $\overline{\mathcal{R}}_0=1$ reduces to the well known threshold when a network with no clustering is considered.

\subsection{Numerical examples}
\begin{figure}[h]
\centering
{\includegraphics[width=0.45\textwidth]{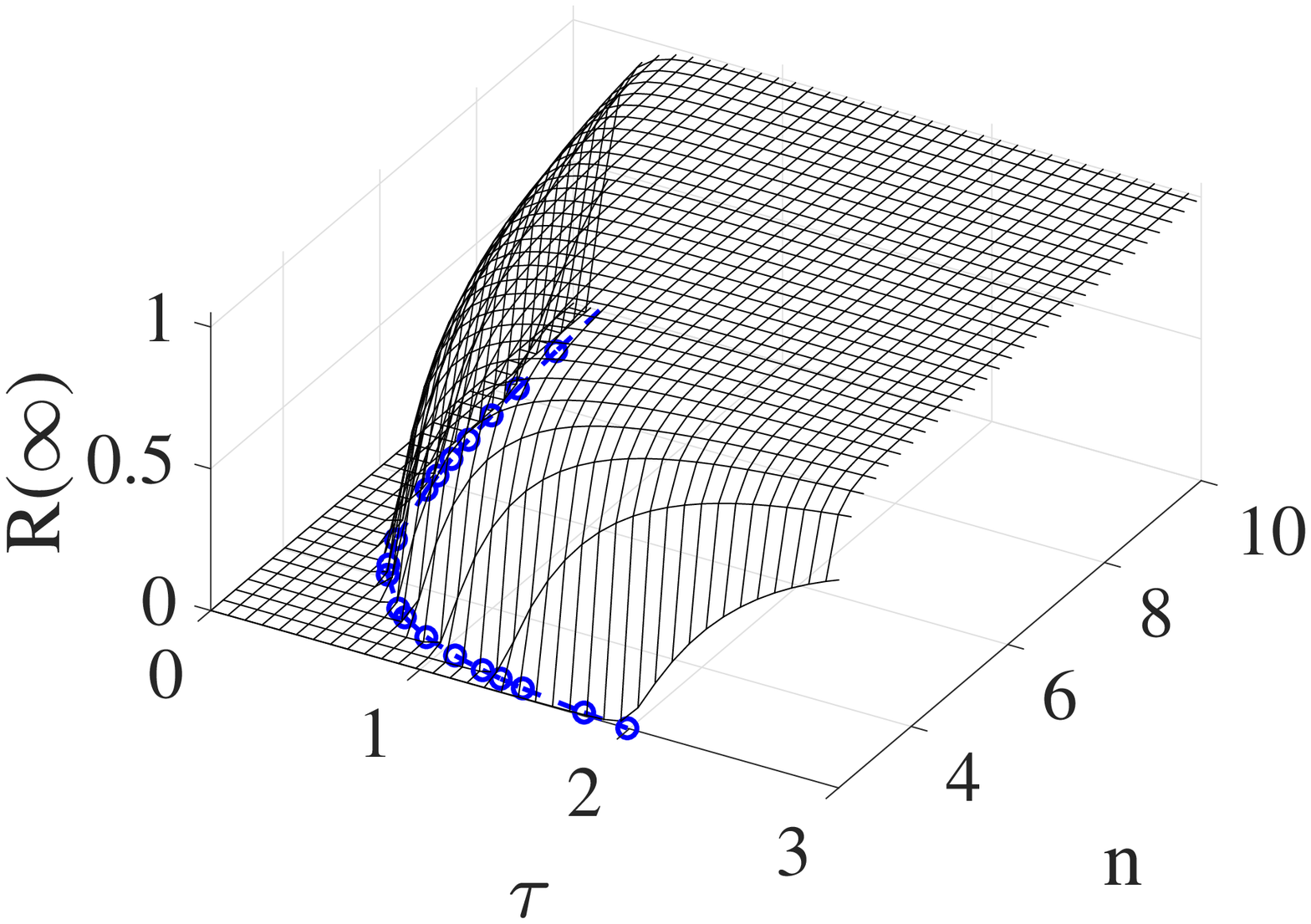}
    }
\hfill
{\includegraphics[width=0.45\textwidth]{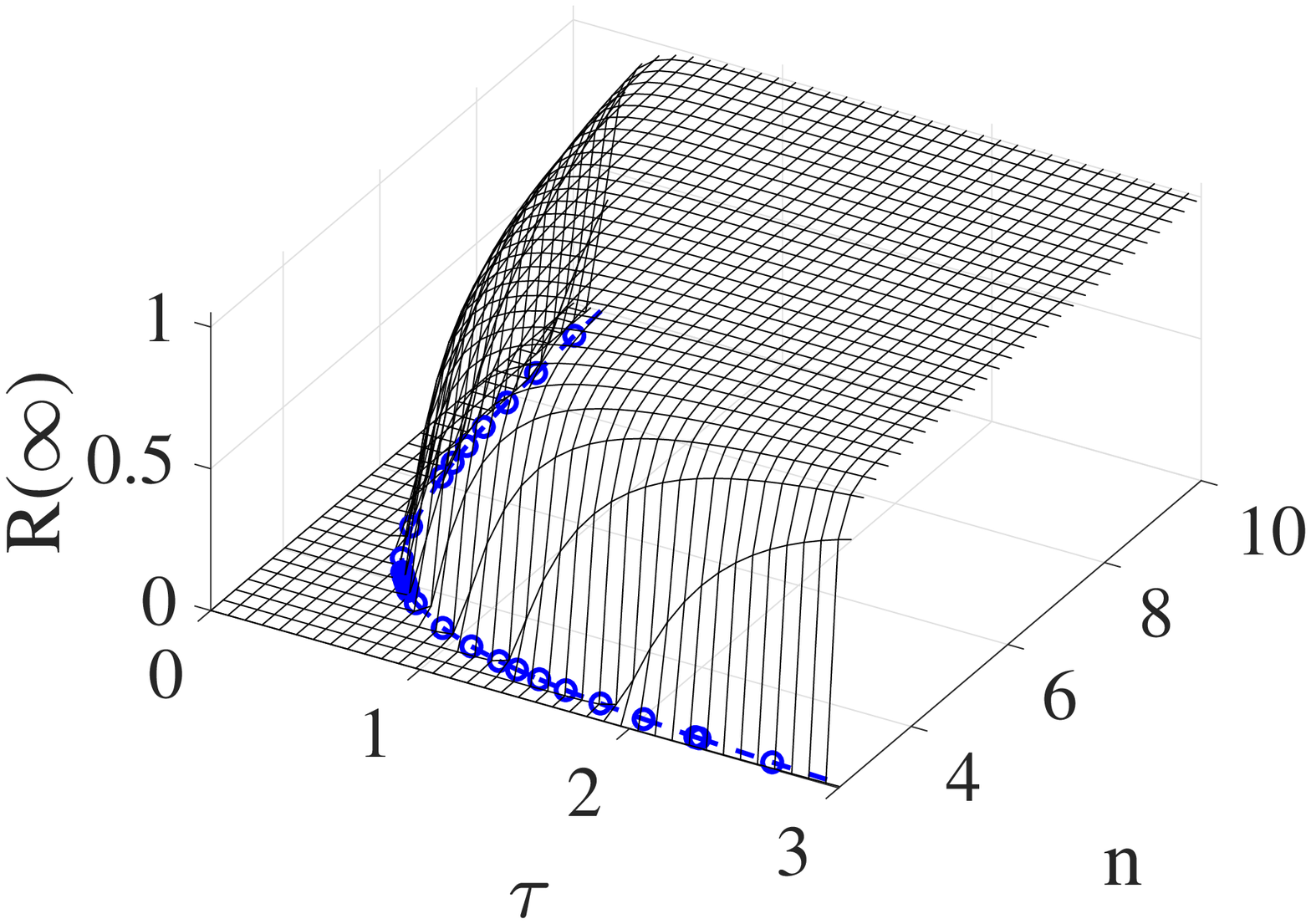}
    }
{\includegraphics[width=0.45\textwidth]{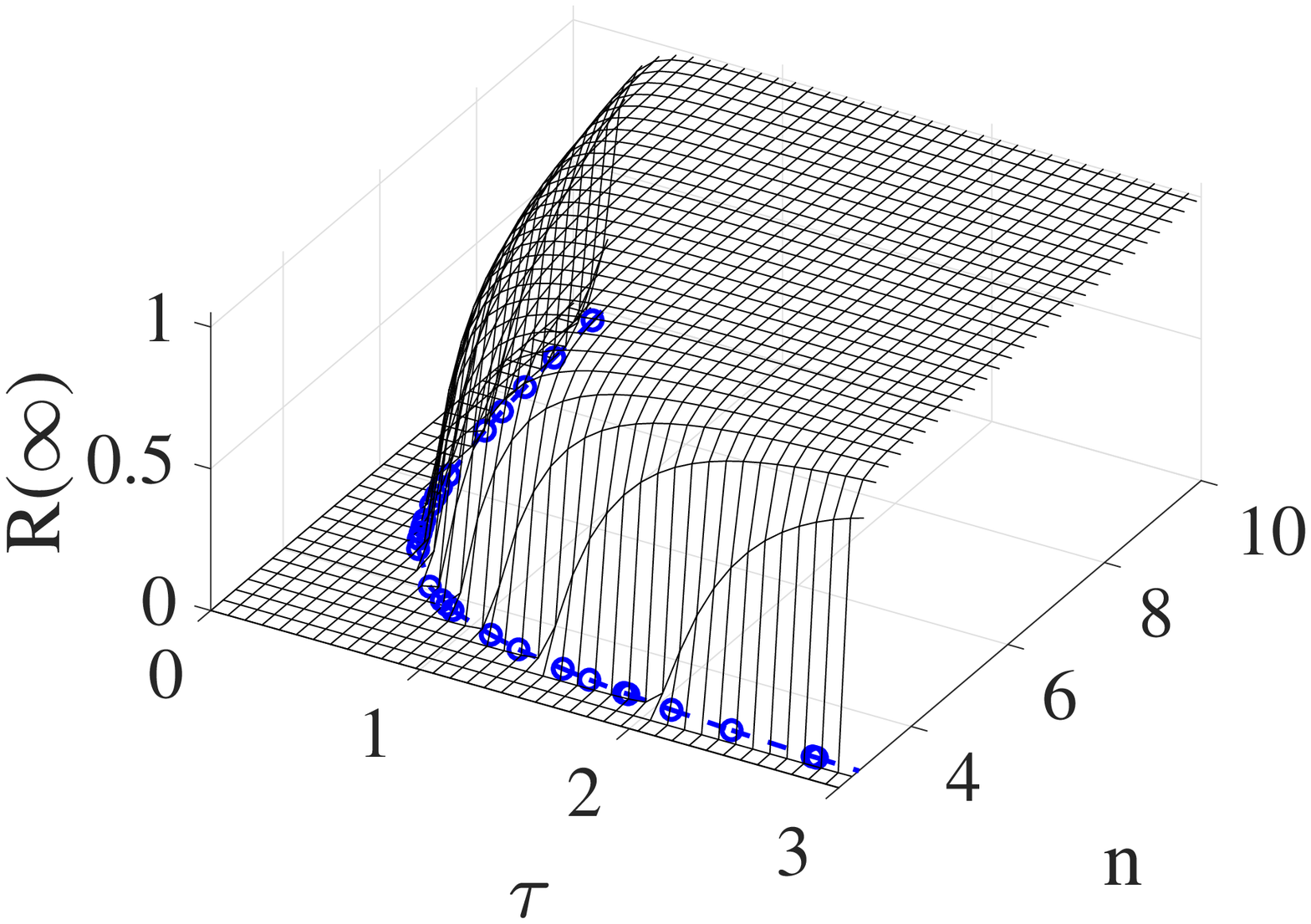}
        }
\hfill
{\includegraphics[width=0.45\textwidth]{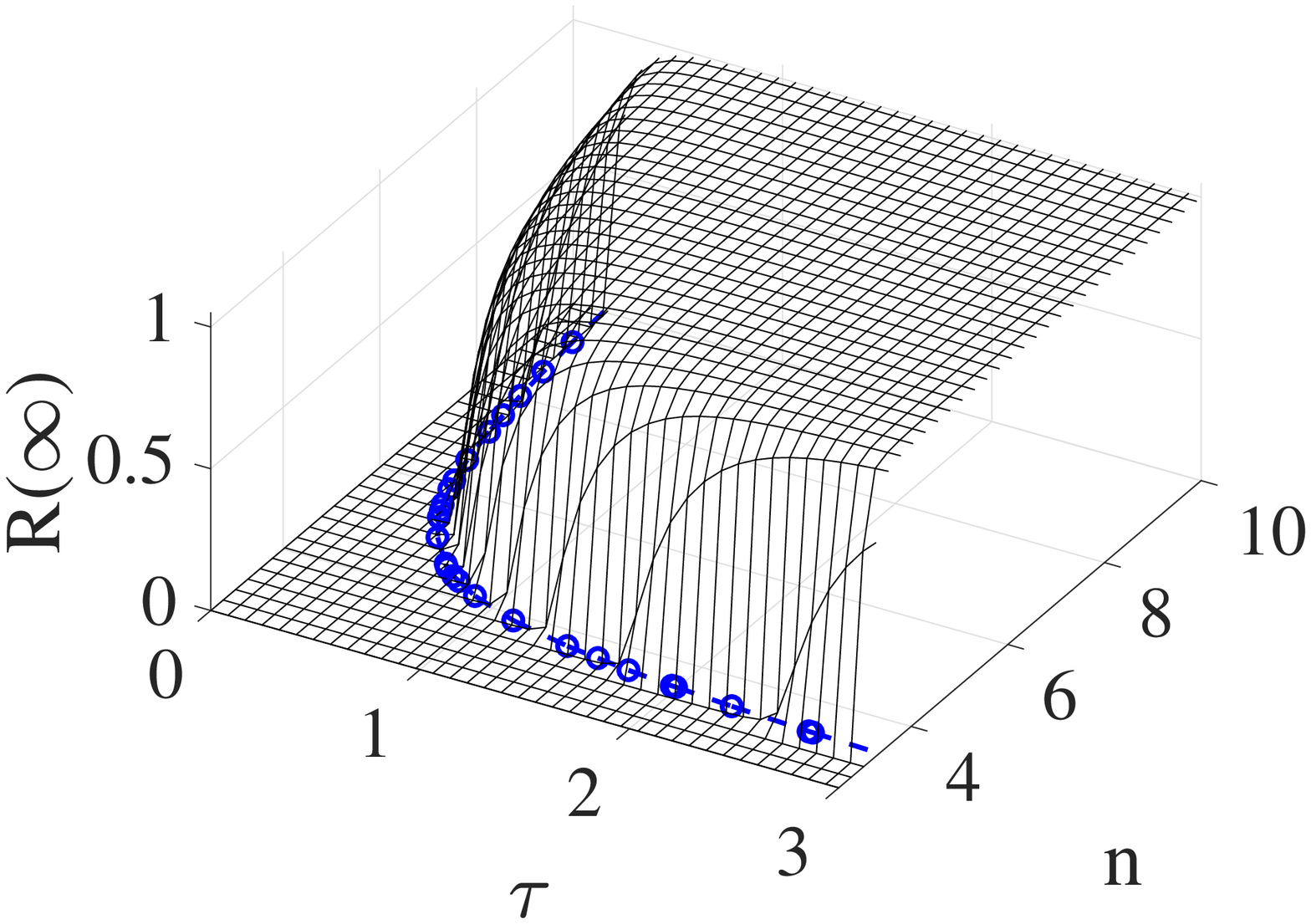}
  }
\caption{Assessing the validity of the epidemic threshold based on the
  asymptotic expansion \eqref{eq:threshold} (dashed line and markers -
  $\circ$) by comparing it to the threshold based on the numerical
  solution of closed pairwise system
  \eqref{eq:dot_S_I_R_imp_close}--\eqref{eq:dot_IR_imp_close}
  (surface). Parameter values are $N=10000$, \ $\gamma=1$ and  the
  clustering coefficients are top
  (from left to right) $\phi=0$, $0.2$, and bottom (from left to
  right) $\phi = 0.4, 0.6$.}
  \label{fig:comp_imp_clo_thres}
\end{figure}
In Fig.~\ref{fig:comp_imp_clo_thres} we show a systematic test of comparing the epidemic threshold generated via solving the closed pairwise system \eqref{eq:dot_S_I_R_imp_close}--\eqref{eq:dot_IR_imp_close} numerically to the epidemic threshold based on the asymptotic expansion \eqref{eq:threshold}, over a wide range of $(\tau,n)$ values. Several observations can be made. First, it is clear that higher values of clustering push the location of threshold to higher $\tau$ and $n$ values, meaning that the limiting effect of clustering on the epidemic spread can only be overcome if either the value of the transmission rate or average degree increases. Second, the agreement between the numerical and asymptotic threshold is excellent for a large range of clustering values. In fact, a slight discrepancy only really seems to appear at around $\phi=0.6$. It is worth noting that finding the final epidemic size numerically can be achieved by using a more compact system. However, the extended system is preferred here since the derivation of the system of fast variables relies upon it.

\section{Discussion}
\label{sec:discussion}

In this paper we set out to obtain an analytic epidemic threshold
using the pairwise model an improved closure to account for
clustering.  This problem has been solved previously in the
unclustered case~\cite{keeling1999effects}.  Here, we went one step further and showed that the quasi-equilibrium can be found as an asymptotic expansion in powers of the clustering coefficient. 
This paper builds on work in~\cite{barnard2018epidemic} and shows that exploiting the presence of fast variables and combining it with perturbation theory leads to a fruitful methodology which allowed us to compute the epidemic threshold analytically from pairwise models with three different closures. Strictly speaking there is no reason why this approach would not apply to other systems with properties similar to those found in the pairwise model. Reflecting on the results in~\cite{barnard2018epidemic} and in the present paper it is obvious that the epidemic threshold is model dependent and care has to be taken if such a model is used to model a real outbreak.

The ODE systems for the fast variables are worth investigating in more detail. We expect that these systems will exhibit a number of steady states. In fact preliminary numerical simulations suggest that the system corresponding to the fast variables~\eqref{eq:alpha}--\eqref{eq:y} has at least one steady state which is identical to the quasi-steady states shown in Fig.~\ref{fig:prev_and_fast_var}. Furthermore, it would be interesting to consider if the idea of fast variables extends to other mean-field models used in epidemiology. In particular it would be worthwhile to investigate if the correlation structure maps onto multi-variable models for heterogenous networks and if this consideration may lead to new insight from existing models. Equally, it remains a challenge to derive compact mean-field models for clustered networks. However, if such models will materialise we expect that our method may be a good candidate when it comes to the analysis of such models. 

Finally, the natural next step would be to test our findings against explicit stochastic network simulations. This was beyond the scope of the present work, whose focus was on exploiting the presence of fast variables and the use of perturbation analysis to determine the epidemic threshold analytically.

\section*{Acknowledgments}

Istv\'an Z. Kiss acknowledges support from the Leverhulme Trust
Research Project Grant (RPG-2017-370). P\'eter L. Simon acknowledges
support from Hungarian Scientific Research Fund, OTKA, (grant
no. 115926).  Joel C. Miller acknowledges support from Global Good.

\bibliographystyle{plain}
\bibliography{sample}
\end{document}